\documentclass{PoS}

\title{Search for $CP$ violation in $D^0\rightarrow \pi^-\pi^+\pi^0$ decays at LHCb}

\ShortTitle{Search for $CP$ violation in $D^0\rightarrow \pi^-\pi^+\pi^0$  decays at LHCb}

\author{\speaker{Shanzhen Chen}\thanks{On behalf of the LHCb collaboration}\\
        University of Manchester\\
        E-mail: \email{shanzhen.chen@cern.ch}}

\abstract{The LHCb experiment has recorded the world's largest sample of charmed meson decays. This paper presents a study of a $D^0$ meson decaying into a final state containing a neutral pion in LHCb. The search for $CP$ violation exploits a novel model-independent unbinned technique to assign a $p$-value for the no $CP$ violation hypothesis. With a data sample size exceeding that of previous measurements by almost an order of magnitude the world's best sensitivity is obtained. 
The $p$-value of no CP violation hypothesis given data of $D^0\rightarrow \pi^-\pi^+\pi^0$  decay analysed is found to be $(2.6 \pm 0.5)\times 10^{-2}$.
}

\FullConference{Flavorful Ways to New Physics - FWNP,\\
		28-31 October 2014\\
		Freudenstadt - Lauterbad, Germany}

\usepackage{amsmath}
\begin{document}

\section{Introduction}
\noindent
The  $D^0\rightarrow \pi^-\pi^+\pi^0$ decay\footnote{$CP$ conjugate is implied throughout this proceeding} is a singly Cabibbo-suppressed neutral $D$ meson decay. 
The amplitudes of this decay channel proceed through $c\rightarrow u\bar{d}d$, 
which lead to the final states  $\rho^0\pi^0$, $\rho^+\pi^-$, and $\rho^-\pi^+$. These final states are common for both $D^0$ and $\bar{D}^0$.

Time-integrated $CP$ asymmetries in $D^0$ decays can have three components: direct
$CP$ violation in decays to specific states, indirect $CP$ violation in $D^0-\bar{D}^0$ mixing, and indirect $CP$ violation in interference of decays with and without mixing. 
 Within the Standard Model, $CP$ violation expected in $D$ decay is of $\textit{O}(10^{-3})$ or less,  
however contributions from particles that are not described in the Standard Model (SM) and participate in the loops of the penguin amplitude can enhance the  $CP$ violation effects expected within the SM \cite{Grossman:2006jg}.
Therefore $CP$ violation in $D^0\rightarrow \pi^-\pi^+\pi^0$ decay provides sensitivity to physics beyond the Standard Model. 

The previous most sensitive studies of the $D^0\rightarrow \pi^-\pi^+\pi^0$ decay has been done by the BaBar~\cite{BaBar} and Belle ~\cite{Belle} collaborations.  At LHCb, with the world's largest dataset, a novel method, known as energy test \cite{energy1, energy2}, is used for the first time to search for $CP$ violation in this decay, and the world's best sensitivity of this decay mode is achieved \cite{LHCb_3pi}. 

\section{Data}
\noindent
Neutral pions reconstructed in LHCb can be grouped into two categories according to the photon cluster structures recorded in the electromagnetic calorimeter ($ECAL$). 
There are pions that have lower momentum (typically p$_T$ $<$ 2 GeV$\//$c) for which both final state photons are reconstructed separately in the $ECAL$ (resolved pions), as well as pions that have higher momentum and thus a merged cluster reconstructed in the $ECAL$ (merged pions). 
The two $\pi^0$ samples provide coverage of complementary regions of the $D^0\rightarrow \pi^-\pi^+\pi^0$ Dalitz phase space and thus the use of both categories of pions contributes significantly to the sensitivity of the analysis.

\begin{figure}[htb]
\begin{center}
                  \includegraphics[width=4.9cm]{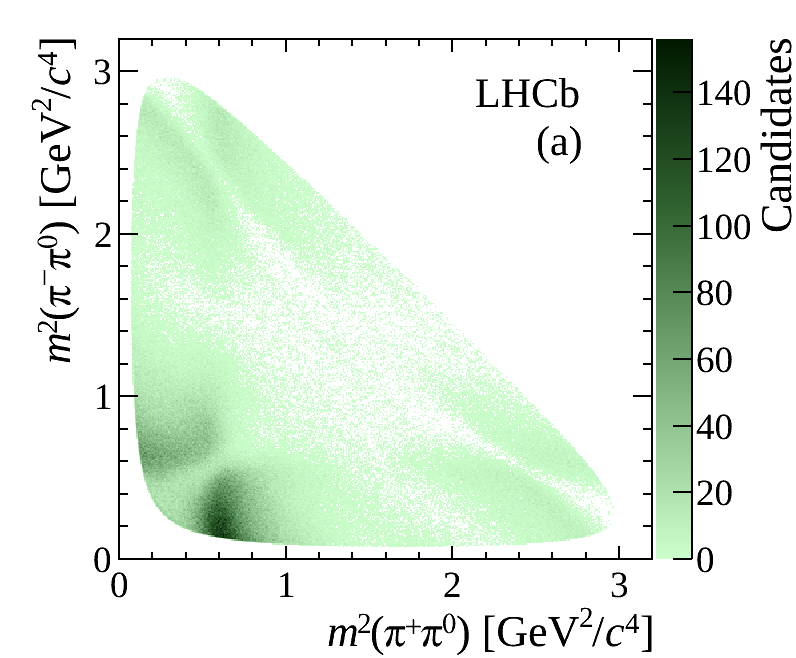} 
                  \includegraphics[width=4.9cm]{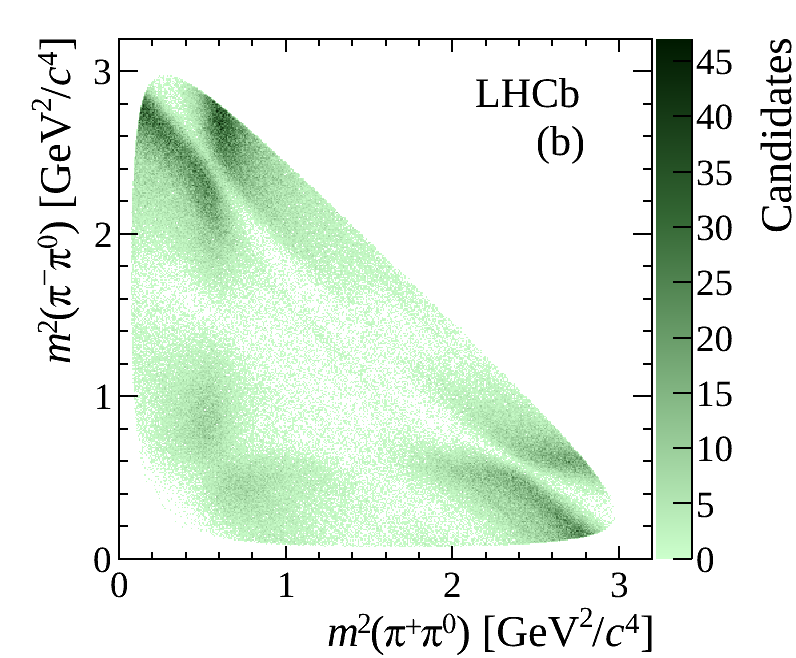} 
                  \includegraphics[width=4.9cm]{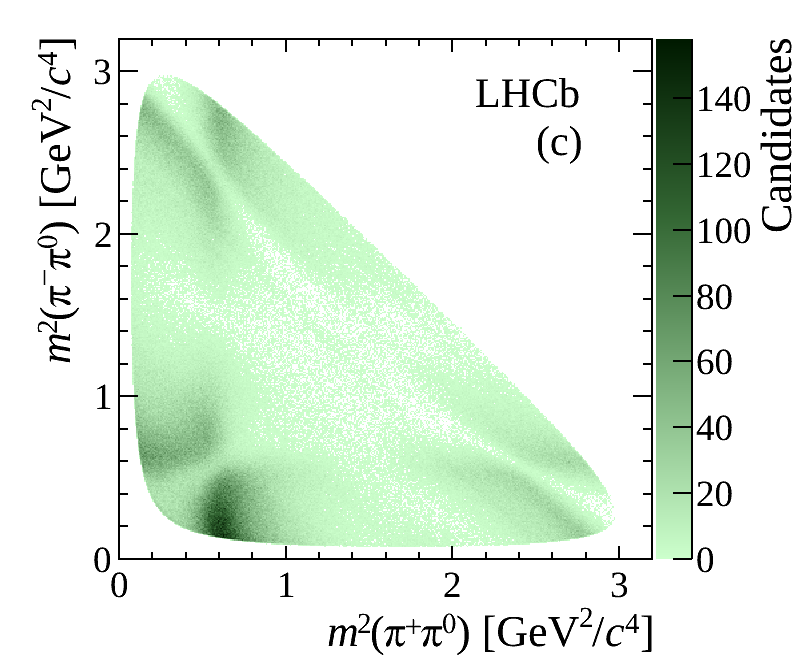}
                  
        \caption{Dalitz plot of the (a) resolved, (b) merged and (c) combined $D^0\rightarrow\pi^-\pi^+\pi^0$ data sample. Enhanced event densities in the phase-space corners originate from the $\rho(770)$ resonances. }
        \label{fig:favortagging}
 \end{center}
 \end{figure}

The data sample used in this analysis is recorded during the 2012 run at a centre-of-mass energy of 8 TeV, with the integrated luminosity $L= 2$fb$^{-1}$. 
The offline selections of the two categories are performed separately. The resolution of merged pions is worse as there is only one cluster in $ECAL$, nevertheless due to large transverse energy and low combinatorial backgrounds, a better purity in merged sample is achieved. After selection, $416\times10^3$ resolved candidates and $247\times10^3$ merged candidates are obtained with purity of $82\%$ and $91\%$, respectively. 
Dalitz plots of resolved and merged samples, and the combination of the two are shown in Fig. \ref{fig:favortagging}.

\section{Energy Test}
\noindent
The energy test is an unbinned model-independent statistical method to search for time-integrated $CP$ violation in multibody decays. 
This method relies on the comparison of the probability density functions of the two flavor conjugated samples in phase space. A test statistic $T$ is formed to realize the comparison, defined as 
\begin{equation}
T\approx\dfrac{1}{n\left(n-1\right)}\sum_{i,j>i}^{n}\psi_{ij}
 + \dfrac{1}{\bar{n}\left(\bar{n}-1\right)}\sum_{i,j>i}^{\bar{n}}\psi_{ij}
 - \dfrac{1}{n\bar{n}}\sum_{i,j}^{n,\bar{n}}\psi_{ij}.
\end{equation}

The metric function $\psi_{ij}\equiv\psi(d_{ij})=e^{-d^2_{ij}/2\sigma^2}$ is chosen to be Gaussian function with
a tunable metric parameter $\sigma$, and $d_{ij}\equiv\vert\Delta\vec{x}_{ij}\vert=\vert(m_{ab}^{2,j}-m_{ab}^{2,i},m_{bc}^{2,j}-m_{bc}^{2,i},m_{ca}^{2,j}-m_{ca}^{2,i})\vert$ is the distance between two events in phase space, where the a, b, c subscripts indicate the final-state particles.  
All three invariant masses are used in the calculation of distances between events.  
Using all three invariant masses does not add information, but it avoids an arbitrary choice that could impact the sensitivity of the method to different $CP$ violation scenarios. 

The three terms in the $T$ formula sum over the weighted distances among $D^0$ decay events, among $\bar{D}^0$ decay events, and between events in different samples, respectively. 
For a $CP$ violating sample, events would be more likely to gather at different regions in the phase space for $CP$ conjugated decays. Therefore, the average distance in each flavour tagged sample would be smaller than the average distance between the two different sets. Since the metric $\psi\left( d_{ij}\right)$ is generally falling with increasing distance, the $CP$ asymmetry would lead to a large $T$ value. 

A permutation method is also used to simulate samples without $CP$ violation by randomly reassigning the flavour of each candidate.
By comparing the nominal $T$ value observed in data to a distribution of $T$ values obtained from permutation samples, a $p$-value under the hypothesis of $CP$ symmetry is obtained as the fraction of permutation $T$ values greater than the nominal $T$ value.

\begin{figure}[tb]
\begin{center}
                  \includegraphics[width=4.9cm]{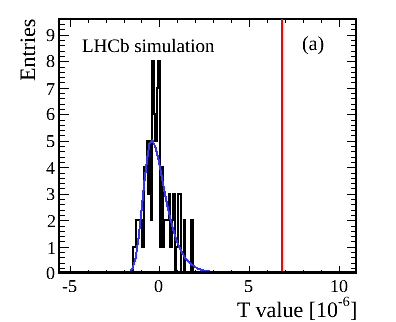} 
                  \includegraphics[width=4.9cm]{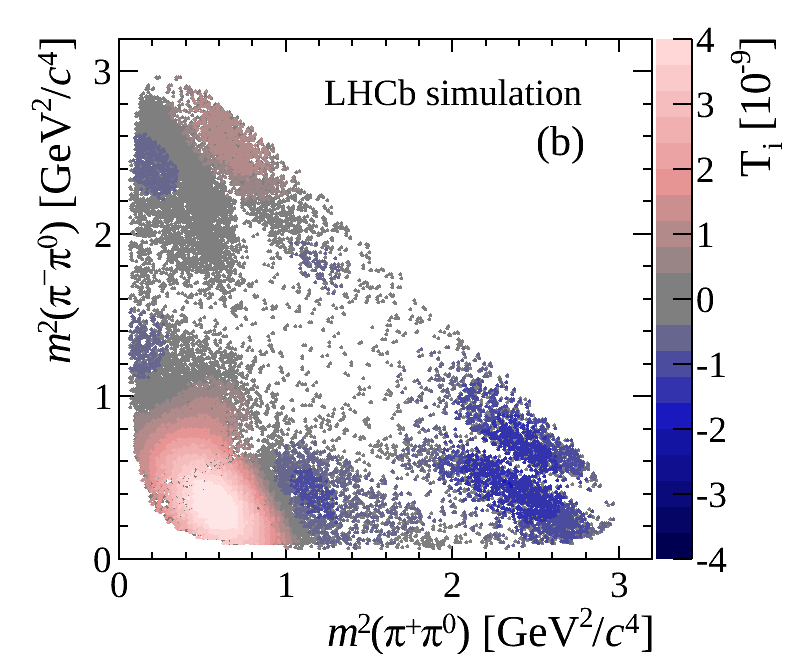} 
                  \includegraphics[width=4.9cm]{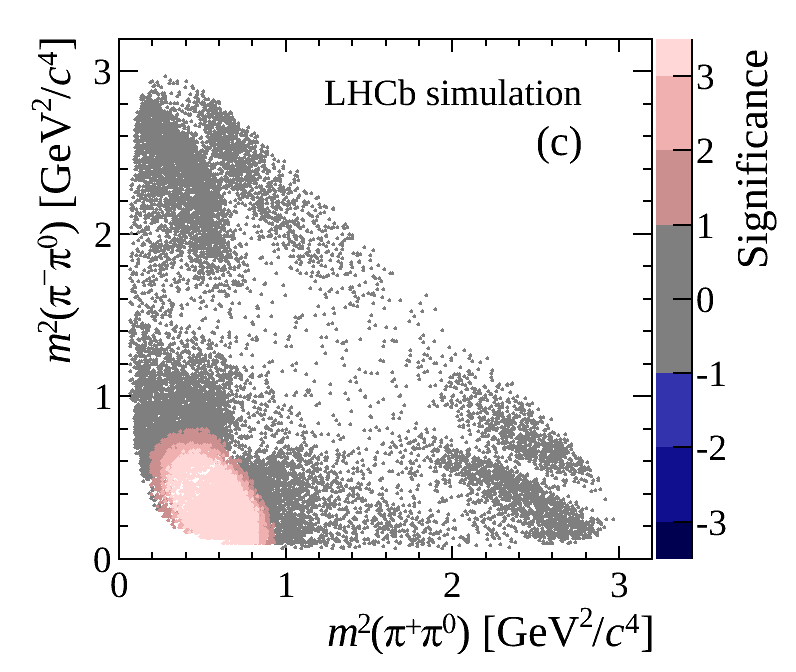}
                  
        \caption{(a) Distribution of permutation $T$ values and nominal $T$ value for a simulated sample with 2\% CP violation in the amplitude of the $\rho^+(770)$ resonance. Permutation $T$s are fitted with a GEV function and the nominal $T$ value is shown as a vertical line. (b) $T_i$ value distributions , and (c) local asymmetry significances are plotted on Dalitz plots for the same simulated sample. }
        \label{fig:mc}
 \end{center}
 \end{figure}

The statistical uncertainty of the $p$-value is obtained as a binomial standard deviation, reflecting the limit number of permutation tests. If large $CP$ violation is observed, the nominal  $T$ value tends to be lying outside the bulk of permutation $T$ values (see Fig. \ref{fig:mc}(a)). In this case, the permutation $T$ distribution can be fitted with a generalised extreme value (GEV) function, which is defined as 
\begin{eqnarray}
f(T;\mu,\delta,\xi) = N \left[1+\xi\left(\frac{T-\mu}{\delta}\right)\right]^{(-1/\xi)-1}&&\nonumber\\
\times\exp\left\{-\left[1+\xi\left(\frac{T-\mu}{\delta}\right)\right]^{-1/\xi}\right\},&&
\end{eqnarray}
with normalisation N, location parameter $\mu$, scale parameter $\delta$, and shape parameter $\xi$.
The $p$-value from the fitted $T$ distribution can be calculated as the fraction of the integral of the function above the nominal $T$ value. This approach is used for the sensitivity studies, which will be described in Sec. \ref{sec:sensitivity}.
The uncertainty on the fitted $p$-value is obtained by randomly resampling the fit parameters within their uncertainties, taking into account their correlations, and by extracting a $p$-value for each of these samplings. 

The contribution of each single event from one flavour, $T_i$, and from the opposite flavour $\bar{T_i}$, to the total $T$ value is defined as
\begin{eqnarray}
\label{eq:Ti}
T_{i} & = & \dfrac{1}{2n\left( n-1\right) }\sum_{j\neq i}^{n}\psi_{ij}-\dfrac{1}{2n\overline{n} }\sum_{j}^{\overline{n}}\psi_{ij},\nonumber\\
\overline{T}_i & = & \dfrac{1}{2\overline{n}\left( \overline{n}-1\right) }\sum_{j\neq i}^{\overline{n}}\psi_{ij}-\dfrac{1}{2n\overline{n} }\sum_{j}^{n}\psi_{ij}.
\end{eqnarray}

A permutation method is also used here to define the level of significance of each event. The visualisation of regions with significant asymmetries is obtained by plotting the asymmetry significance in terms of each event's contribution on Dalitz plot (Fig. \ref{fig:mc}(b)(c)).  

The practical limitation of this method is that the number of mathematical operations scales quadratically with the sample size. Furthermore, a significant number of permutations is required to get a sufficient precision of the $p$-value. In this analysis, graphics processing units (GPUs) \cite{GPU} are used in the energy test calculation. The parallel architecture of GPUs reduces calculation time by about 95\% in this analysis. 

\section{Sensitivity studies}
\label{sec:sensitivity}

\begin{table}[t]
\caption{Overview of sensitivities to various $CP$ violation scenarios.
$\Delta A$ and $\Delta \phi$ denote, respectively,
change in amplitude and phase of the resonance $R$.}
\centering
\begin{tabular}{lll}
\hline \hline
$R$ ($\Delta A$, $\Delta \phi$) & $p-$value (fit) \\
\hline
$\rho^0$ $(4\%$, $0^\circ)$ & $3.3^{+1.1}_{-3.3}\times10^{-4}$  \\
$\rho^0$ $(0\%$, $3^\circ)$ & $1.5^{+1.7}_{-1.4}\times10^{-3}$   \\
$\rho^+$ $(2\%$, $0^\circ)$ & $5.0^{+8.8}_{-3.8}\times10^{-6}$   \\
$\rho^+$ $(0\%$, $1^\circ)$ & $6.3^{+5.5}_{-3.3}\times10^{-4}$  \\
$\rho^-$ $(2\%$, $0^\circ)$ & $2.0^{+1.3}_{-0.9}\times10^{-3}$   \\
$\rho^-$ $(0\%$, $1.5^\circ)$ & $8.9^{+22}_{-6.7}\times10^{-7}$   \\
\hline \hline
\end{tabular}
\label{tab:sensitivity_overview}
\end{table}

\noindent
A study of sensitivity with different types and amounts of asymmetries is required in the interpretation of results. This sensitivity study is done with simplified Monte Carlo data generated according to the model described in Ref. \cite{BaBar} with the generator package Laura++ \cite{Laura++}.

The selection efficiency is also taken into account in the sensitivity study as it varies strongly across phase space. The selection efficiency is obtained using full LHCb detector simulation.  
The sensitivity studies are performed using a cocktail of simulated signal and background events. The signal sample is generated using Laura++ taking into account the Dalitz plot efficiency model, while the background sample is generated according to the sidebands distribution.

A study of the metric parameter $\sigma$ is also performed in the sensitivity studies. $\sigma$ plays the role of bin size in a binned method. It should be larger than the resolution as otherwise asymmetries would be washed out by fluctuations within the scale of the metric. At the same time it should be small enough to resolve the regions where asymmetries lead to different densities. Within these two constraints, various values of $\sigma$ are tested and $\sigma = 0.3~$GeV$^2/$c$^4$ is chosen as the default value.

With 100 permutations, the studies of different $CP$ asymmetry scenarios and their sensitivities are reported in Table \ref{tab:sensitivity_overview}.

\section{Results and conclusions}
\begin{figure}[tb!]
\begin{center}
                  \includegraphics[width=5.5cm]{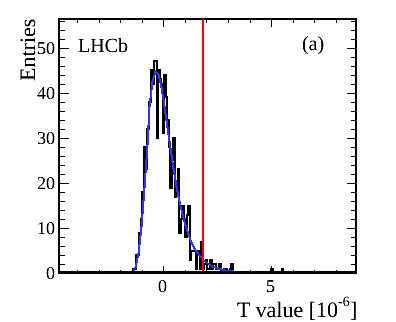} 
                  \includegraphics[width=5.5cm]{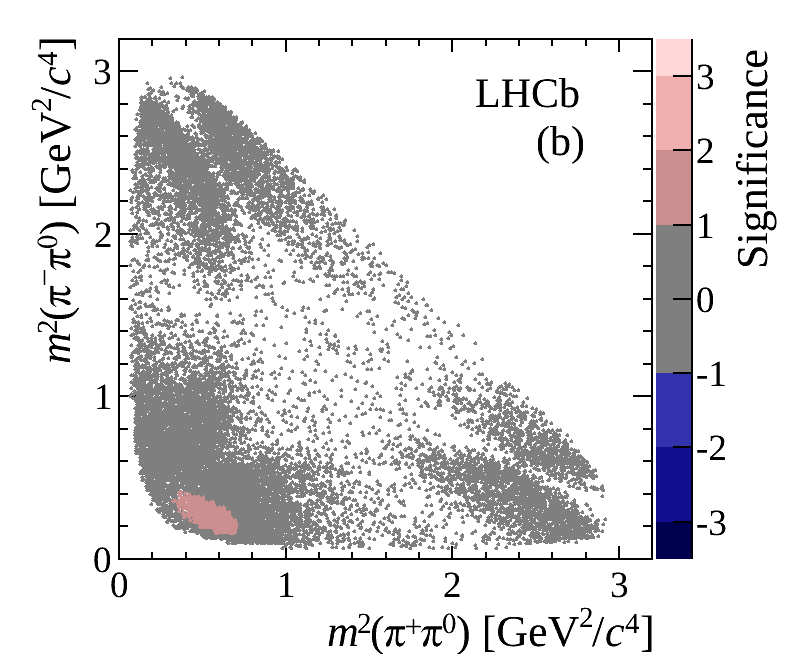} 
                  
        \caption{(a) Permutation (no $CP$ asymmetry) $T$ value distribution showing the fit function and the measured $T$ value as a vertical line. (b) Visualisation of local asymmetry significances. The positive (negative) asymmetry significance is set for the $D^0$ candidates having positive (negative) contribution to the measured $T$ value, respectively. }
        \label{fig:results}
 \end{center}
 \end{figure}

\noindent
By counting the fraction of permutation $T$ values above the nominal $T$ value with 1000 permutations, a $p$-value of $(2.6\pm0.5)\times10^{-2}$ for default $\sigma$ ($0.3GeV^2/c^4$) is obtained. Varying the metric parameter results in the $p$-values listed in Table \ref{tab:results}. All results are consistent with the no $CP$ violation hypothesis. The nominal $T$ value for data and permutation $T$ values for no $CP$ asymmetry hypothesis, and the visualisation of significant regions on Dalitz plot are shown in Fig. \ref{fig:results}. 

In summary, a search for CP violation in the $D^0\rightarrow\pi^-\pi^+\pi^0$ decay is performed using a novel unbinned model-independent technique, the world's best sensitivity of this decay from a single experiment is achieved. The data are found to be consistent with the no $CP$ violation hypothesis with a $p$-value of $(2.6\pm0.5)\times10^{-2}$.

\begin{table}[t!]
\caption{Results for various metric parameter values.
The $p$-values are obtained with the counting method. All p-values are found to be consistent with no $CP$ violation hypothesis.}
\centering
\begin{tabular}{cc}
\hline \hline
$\sigma$~[GeV$^2/$c$^4$ ] & $p-$value \\
\hline
0.2 & $(4.6 \pm 0.6)\times10^{-2}$ \\
0.3 & $(2.6 \pm 0.5)\times10^{-2}$ \\
0.4 & $(1.7 \pm 0.4)\times10^{-2}$ \\
0.5 & $(2.1 \pm 0.5)\times10^{-2}$ \\
\hline \hline
\end{tabular}
\label{tab:results}
\end{table}

\end{document}